%% file: root.tex
\documentclass[letterpaper, 10 pt, conference]{ieeeconf}  % Comment this line out if you need a4paper

\IEEEoverridecommandlockouts                              % This command is only needed if 
\usepackage[pdftex]{graphicx}
\usepackage{caption}
\usepackage{subcaption}
\usepackage{siunitx}
\usepackage{algpseudocode}
\usepackage{xspace}

\usepackage{multirow}

\usepackage[table,x11names]{xcolor}
\usepackage{cases}
\usepackage{array}
\usepackage{color}
\definecolor{pink}{rgb}{1,0,1} %{.5,0,0} %
\usepackage{url}
\usepackage{float}

\newcommand \ignore[1]{}

\title{\LARGE \bf
Continuous Authentication of Wearable Device Users from \\Heart Rate, Gait, and Breathing Data
}

\author{%
William~Cheung and Sudip Vhaduri\\
Fordham University, Bronx, NY 10458, USA\\
\{wcheung5, svhaduri\}@fordham.edu\\
}

\begin{document}

\maketitle
\thispagestyle{empty}
\pagestyle{empty}

%%%%%%%%%%%%%%%%%%%%%%%%%%%%%%%%%%%%%%%%%%%%%%%%%%%%%%%%%%%%%%%%%%%%%%%%%%%%%%%%
\begin{abstract}
  The security of private information is becoming the bedrock of an increasingly digitized society. While the users are flooded with passwords and PINs, these gold-standard explicit authentications are becoming less popular and valuable. Recent biometric-based authentication methods, such as facial or finger recognition, are getting popular due to their higher  accuracy. However, these hard-biometric-based systems require dedicated devices with powerful sensors and authentication models, which are often limited to most of the market wearables. Still, market wearables are collecting various private information of a user and are becoming an integral part of life: accessing cars, bank accounts, etc. Therefore, time demands a burden-free implicit authentication mechanism for wearables using the less-informative soft-biometric data that are easily obtainable from modern market wearables. 
  In this work, we present a context-dependent soft-biometric-based authentication system for wearables devices using heart rate, gait, and breathing audio signals. 
  From our detailed analysis using the ``leave-one-out'' validation, we find that a lighter $k$-Nearest Neighbor ($k$-NN) model with $k = 2$ can obtain an average accuracy of $0.93 \pm 0.06$, $F_1$ score $0.93 \pm 0.03$, and {\em false positive rate} (FPR) below $0.08$ at 50\% level of confidence, which shows the promise of this work.
\end{abstract}

\input{introduction}
\input{approach}

\input{authentication}

\input{discussion}

\bibliographystyle{IEEEtran}
\bibliography{reference}

\end{document}

%% file: introduction.tex
\section{Introduction}\label{introduction}

\subsection{Motivation}
Internet of Things (IoT) have increasing access to a multitude of devices with advanced capabilities that allow us to remotely collect information or control physical objects.\ignore{Various examples of such systems include alarm systems, entertainment devices, vehicles, and smart home devices, to name a few. IoT systems are connected to existing network infrastructure providing new interfaces that integrate the physical world with computer-based systems.} Along with the growth of IoT, smartphones and wearables have advanced in their sensing and computational capabilities to a point which enable many new applications and usage scenarios to emerge~\cite{vhaduri2020adherence,vhaduri2020nocturnal,vhaduri2019nocturnal,vhaduri2019towards,al2009load}.\ignore{ While smartphones are already widely used \cite{seneviratne2017survey},} \ignore{On the other hand, wearables' \textcolor{pink}{security is} relatively new and still growing with the arrival of new applications.} Some include the ability to identify a user to third party services \cite{bianchi2016wearable}, store sensitive user information (i.e., passwords, credit card information) \cite{nguyen2017smartwatches}, manage financial payments \cite{seneviratne2017survey}, access phones and other paired devices \cite{kumar2016authenticating}, unlock vehicles \cite{nguyen2017smartwatches}, monitor or track individuals (e.g., child monitoring or fall detection), and assess an individual's health and fitness. According to a recent market report, a 72.7\% increase in wearable shipments and an associated increase in sales revenue of 78.1\% are predicted from 2016 to 2022~\cite{wearable_shipments_revenue}. 

However, wearables also raise new challenges, especially in terms of security. Unauthorized access to a wearable can enable imposters to steal information from other sensitive IoT objects, which poses a significant risk \cite{zeng2017wearia}.\ignore{Unauthorized users could also steal data on the wearables itself, e.g., many applications and services provided by a wearable depend on sensor and user data stored on the device to grant access.} \ignore{Another concern is the reliability and accuracy of the physiological and activity data collected by wearables. Many healthcare providers and researchers rely on wearables to monitor their patients or study participants remotely and an intentional device sharing between target and non-target users might lead to inaccurate and faulty assessment.} 
%\textcolor{cyan}{For example health care provider's tolerance of misclassification especially false negatives of patient identification is very low and open to drug abuse.} 
Illustrations include, the stealing of sensitive personal information or delivering faulty dosage of a drug to a remote patient. 
Therefore, there is an imperative need for a robust and accurate authentication mechanism specifically for wearable device users.
Existing wearable devices either have no authentication systems or authentication mechanisms that are often knowledge-based regular PIN locks or pattern locks~\cite{nguyen2017smartwatches}, which suffer from scalability issues~\cite{unar2014review}. 
\ignore{With an increasing reliance on protected devices, a user can be overwhelmed with passwords or PIN requests to obtain access to various data and services. Knowledge-based approaches also require user interactions with the display, which may either be inconvenient to certain class of users or even completely absent in many wearables~\cite{unar2014review,zeng2017wearia}.}
Many times, users opt to completely disable security mechanisms out of convenience, as the design hinders the implementation of security itself. %\textcolor{cyan}{However an implicit system would not require user intervention increasing ease of use, enabling more security systems.}
However, an implicit and continuous authentication system does not require any explicit user input, and thereby, such an authentication can work seamlessly without imposing any user burden.

\subsection{Related Work}

\subsubsection{Wearable Constraints}
Wearable device user authentication is a relatively new field of research compared to mobile authentication~\cite{bianchi2016wearable,blasco2016survey,cornelius2012wears,unar2014review}.
The limited display sizes of wearables add another constraint that limits the choices of authentication mechanisms~\cite{bianchi2016wearable,vhaduri2017wearable}. But as technology advances, companies such as Samsung, Fitbit, Apple, Garmin, and Embrace can provide lower level granularity in data. In addition to this, increasingly more biometrics are available as more sensors are being added such as microphone, electrocardiograms (ECG), and GPS. 
However, some of the newer sensors, such as ECG are not yet accurate enough\ignore{up to security or medical standards}. 
For example, researchers have found that, although for people over the age of 85 Apple accurately detects atrial fibrillation at a rate of 96\%, for people under 55\textcolor{pink}{,} it only correctly diagnoses atrial fibrillation 19.6\% of the time~\cite{Apple_ECG}. 
\ignore{However, given better ECG technology, implicit authentication models will have a better performance.} 
Another group of researchers designed wrist strapped ECG reader and developed an authentication system with an accuracy of 93.5\%, which is limited by the ease of use and placement issues~\cite{yan2019towards}. Therefore, an authentication scheme that can utilize data from different sensors, such as photo-plethysmography (PPG), accelerometer, gyroscope, and microphone, which are readily available on market wearables, could be more realistic to develop a non-stop implicit wearable authentication system. 

\subsubsection{Multi-modal Biometric Authentication}
In previous work, combinations of biometrics were used to form multi-modal biometric authentication systems for increased reliability compared to unimodal systems, which often suffer from noisy data, intra-class variations, inter-class similarities, and spoof attacks~\cite{ghayoumi2015review}. For multi-modal authentication systems, researchers have utilized different hard- and soft-biometrics. However, due to accuracy concerns in sensing capabilities and relatively low computational power of wearables, these multi-modal approaches are typically not implemented for implicit and continuous authentication on state-of-the-art wearables.

\subsubsection{Wearable Authentication}
 Researchers recently proposed authentication techniques that are more suitable for wearables, focusing more on approaches based on {\em behavioral biometrics}, such as gait~\cite{al2017unobtrusive,cola2016gait,johnston2015smartwatch}, activity types~\cite{bianchi2016wearable,zeng2017wearia}, gesture~\cite{davidson2016smartwatch}, keystroke dynamics~\cite{acar2018waca} and {\em physiological biometrics}, such as PPG signals~\cite{karimian2017non}. Almost all of these studies are based on project specific generated datasets\ignore{ databases and the accuracy of these techniques is often verified with limited numbers of subjects and over short time periods}. All of these user authentication techniques are limited in the scope of use, e.g., gait-based behavioral authentication approaches~\cite{cola2016gait,johnston2015smartwatch} only work during walking. While other projects have addressed some of the limitations of gait-based approaches by considering different types of gestures~\cite{davidson2016smartwatch} or activities~\cite{bianchi2016wearable,zeng2017wearia}. All of these models are based on movement and thereby, fail to work in the very common human state of being sedentary~\cite{acar2018waca,vhaduri2017wearable}. Authentication approaches using physiological biometric data, such as ECG and bioimpedance~\cite{cornelius2012wears} require very fine-grained samples and sensor readings are easily affected by noise, motion, etc., but these biometrics are always available. Focusing on one or a particular set of biometrics restricts the usability of a continuous authentication model, but with a collection of various combinations it is possible to build a more robust authentication system. \ignore{This method should be able continue and adjust when certain biometrics become unavailable or available.}

\subsection{Contributions}
The main contribution of this paper is an exploration of implicit and non-stop user authentication schemes utilizing multiple biometrics. 
In this approach, we use three different types of soft-biometrics: heart rate, gait, and breathing all of which are easily obtainable on most of the market wearables.
\ignore{Compared to previous work~\cite{vhaduri2017wearable,vhaduri2019multi}, where we use hybrid-biometrics, such as calorie burn that can be affected by a user's self-reported input, such as age, height, and weight, in this work we focus on the different biometrics that can be measured without the user's self-reported input.}
While minute-level coarse-grained heart rate samples can be less informative, they are available almost all the time when a user wears the device. Adding another biometric, such as gait, can improve the performance. However, unlike the heart rate data, gait is available only when a user walks. Compared to gait, breathing audio signal can have a better availability, but it also suffers from various issues, such as a user's distance from the microphone, presence of other sounds. Therefore, it is important to develop an authentication approach that can combine multiple biometrics based on contexts or scenarios, e.g., availability of biometrics, and can be easy to implement on wearables\ignore{ and can have a way to decide which modalities should be used based on contexts or scenarios}. In this work, we present a multi-biometric-based context-driven approach that works both in {\em sedentary} and {\em non-sedentary} periods (Section~\ref{approach}). From our detailed feature computation and selection, hyper-parameter optimization, and finally, modeling with different classifiers, we are able to authenticate a user with an average accuracy of $0.93 \pm 0.06$, $F_1$ score $0.93 \pm 0.03$ (Section~\ref{model_performance}), and FPR below $0.08$ at 50\% level of confidence while classifying (Section~\ref{accessing_error_rate}). 

%% file: approach.tex
\section{Approach}\label{approach}

In this paper, we intend to demonstrate the importance and effectiveness of different biometrics to identify wearable-users using machine learning modeling. Before we describe the details of the analysis, we first introduce the datasets, pre-processing steps, feature engineering, and methods used in this work.

\subsection{Datasets}\label{datasets}
%% Will is going to write down brief description of different datasets along with a citation/reference to either their papers or dataset URL (if paper doesn't exist)
We use three different datasets in our analysis. 

\begin{itemize}
    \item Fitbit dataset: We use the heart rate data collected at a rate of one sample per minute using the Fitbit Charge HR device from three subjects similar to our previous work~\cite{vhaduri2016assessing,vhaduri2016cooperative,vhaduri2017discovering,vhaduri2017wearable,vhaduri2017towards,vhaduri2018hierarchical,vhaduri2018biometric,vhaduri2018impact,vhaduri2018opportunisticICHI,vhaduri2018opportunisticTBD,vhaduri2016design,vhaduri2016human,vhaduri2017design,vhaduri2019multi,chihyou2020estimating}. 
    Data was collected during various activity levels ranging from sedentary to high activity.
    \item Gait dataset: We use the WISDM dataset~\cite{WISDM}, where three (i.e., x, y, z directions) linear and angular acceleration readings were collected from  accelerometer and gyroscope, respectively, at a rate of one sample in 50 ms from the LG G Watch with Wear 1.5 operating system. Data was collected under normal walking condition.
    \item Audio dataset: In this work, we use the breathing audio clips obtained from the ESC-50 dataset~\cite{ESC50}, where audio clips were, sampled at 22.05 kHz, around 5 seconds long. Data was collected, holding a device close to a subject during an idle state.
\end{itemize}

\subsection{Data Pre-Processing}\label{pre-proc}

Since we are using real-world datasets, we need to process the dataset before using it. 
First, we need to segment the continuous stream of biometrics, such as heart rate, gait information, and desired audio events (i.e., breathing). 
Then, we perform data augmentation to simulate more physiological states from the collected data to increase robustness of experiment.
Finally, we compute and select influential features before constructing authentication models. 

\subsubsection{Data Segmentation}\label{vData}
Since heart rate and gait data were sampled at different frequencies, we segment the heart rate and gait samples into 10-sample windows\ignore{ in order to synchronize the two time-series data and} to obtain stable and rich information\ignore{ before feature generation}.
Unlike the heart rate or gait data, the audio data comes with other types of sounds in addition to desired breathing sounds. Additionally, some clips come with multiple breathing events separated by silence or noisy parts. Therefore, we segment the ESC-50 audio clips to fetch single inhalation breathing events. Thereby, we obtain around six inhalation breathing events per subject. 
 \ignore{Per 5 second clip of breathing audio sound,} 
% Per audio clip, each inhalation breath was taken as an individual audio event with a total of six events per subject. 
 \ignore{The rest of the audio recording that did not satisfy this parameter was not used in this study.}
%A similar filtering approach is applied for calorie burn and MET. For our analysis, we consider data from 421 Fitbit users.  

\subsubsection{Audio Data Augmentation}
%\subsubsection{Data Augmentation}
\ignore{Sampling size of some aspect of the data sources were limited and a need to generate simulated estimates of other data was needed in order to facilitated better testing.} 
%Realistic augmentation of the original data is done with the audio biometrics to cope with limited data.  
\ignore{Fingerprints are inherently static but the types of imaging might not be. In order to simulate this, we augmented the original data points. For images changes with various rotations, zooms, and combinations rotation with zooms were used, while for audio pitch and speed adjustments were made.}
Breathing audio could be altered due to change of environments, physical state, or mood. To simulate this and capture the variations, we augment the original audio breathing events using various pitch shifts and speed changes.
\begin{itemize}
    \item Pitch shift: We consider 15 different pitch shifts ranging from -$\frac{7}{2}$ to $\frac{7}{2}$ with $\frac{1}{2}$ increments
    \item Speed change: We consider seven speed changes ranging from .25x to 2x times the speed of an original clip with an increment of .25x, skipping 1x since that would represent the original clip, which we have already included as a pitch shift with value $0$.
\end{itemize}
Thereby, each original breathing clip is augmented using a total of 22 modifications.

%\subsubsection{Data Synchronization}
%In this work, we concatenate the heart rate and gait data into 10-sample windows to compute features. 
%Since heart rate and gait data were sampled at different frequencies, therefore, we segment the heart rate and gait samples into \textcolor{red}{\bf 10-sample} windows in order to synchronize the two time-series data and obtain stable and rich information before feature generation.

\ignore{----------------------
\begin{figure}[ht]
\centering
\scalebox{.3}{\includegraphics{figNew/"HG_SelectFromModel_(3)".png}}
\caption{Top 21 heart rate and gait features selected (top 14 green bars are used for modeling) using the SelectFromModel approach with the explanation threshold, $p = 0.90$.}
\label{HG SelectFromModel}
\end{figure}
----------------------}

\begin{table*}[!t]
\caption{Summary of features selected from different biometrics}
\label{Best_Features}
\centering
\begin{tabular}{l|l|p{8cm}} %% Number of Columns 
\hline
Biometrics & Selector (parameters) & Selected features \\
\hline
Heart rate & SelectKBest ($K$ = 10) & $p25$, $\mu$, $rss$, $rms$, $max$, $Mdn$, $p75$, $\kappa$, $\gamma$, $P$\\
\hline 
Heart rate and gait & SelectFromModel (p = 0.90) & Z-gy $\kappa$, Y-gy $\kappa$, Y-acc $\mu$, $\mu$, Z-gy $\sigma^2$, Z-acc $\kappa$, X-acc $\kappa$, Y-gy $\sigma^2$, Y-acc $\kappa$, X-acc $\mu$, X-acc $\sigma^2$, Z-acc $\sigma^2$, Y-acc $\sigma^2$ \\
\hline
Heart rate and breathing & SelectKBest ($K$ = 10) & MFCC3, MFCC7, MFCC4, MFCC6, MFCC9, MFCC11, MFCC15, MFCC38, MFCC13, MFCC40\\
\hline
\end{tabular}
\end{table*}

\begin{table*}[!t]
\caption{Summary of features selected from heart rate, gait, and breathing biometrics together}
\label{HGS_Features}
\centering
\begin{tabular}{l|p{12cm}} %% Number of Columns 
\hline
Selector (parameters) & Selected features\\
\hline
%Heart Rate and Gait and Breathing Model\\
%\hline
SelectFromModel (p = 0.90) & MFCC3, MFCC4, MFCC7, MFCC1, MFCC6, Z-gy $\kappa$,  MFCC9, Y-acc $\mu$, MFCC13, Y-gy $\kappa$, Z-acc $\kappa$, MFCC10, MFCC15, X-acc $\kappa$, MFCC11, MFCC18, MFCC36, MFCC38, MFCC26, MFCC17, MFCC12, MFCC14, Z-gy $\sigma^2$, $\mu$,\\
\hline
\rowcolor{lightgray} SelectKBest ($K$ = 10) & MFCC3, MFCC7, MFCC4, MFCC6, MFCC1, Y-acc $\mu$, MFCC9, MFCC38, MFCC2, MFCC11\\
\hline
\end{tabular}
\end{table*}

%\subsection{Feature Engineering}\label{f_eng}

%\subsubsection{Feature Computation}\label{f_comp}
\subsection{Feature Computation}\label{f_comp}

%In this work, we concatenate the heart rate and gait data into 10-sample windows to compute features. 

%% Will has to check this list for heart rate if he hasn't used any of these features; he also has to add definition of features computed from other data, such as accel/gyroscope, audio, finger image with appropriate citation/references to papers/urls
We compute the following sets of candidate features.
\begin{itemize}
    \item Heart rate features: From the windows of 10 samples we compute 21 statistical features: mean ($\mu$), median ($Mdn$), standard deviation ($\sigma$), variance ($\sigma^2$), coefficient of variance ($cov$), range ($ran$), coefficient of range ($coran$), first quartile or $25^{th}$ percent ($p25$), third quartile or $75^{th}$ percent ($p75$), max ($max$), inter quartile range ($iqr$), coefficient of inter quartile ($coi$), mean absolute deviation ($mad\_Mdn$), median absolute deviation ($mad\_\mu$), energy ($E$), power ($P$), root mean square ($rms$), root sum of squares ($rss$), signal to noise ratio ($snr$), skewness ($\gamma$), and kurtosis ($\kappa$), described in~\cite{vhaduri2019multi}.
    \item Gait features: We compute mean ($\mu$), variance ($\sigma^2$), and kurtosis ($\kappa$) from each window of x-, y-, and z-axis readings obtained from gyroscope and accelerometer, separately.
    \item Audio features: From each inhalation breathing event we compute 40 Mel-frequency cepstral coefficients (MFCCs), where, $MFCCi$ represents the $i^{th}$ coefficient.
\end{itemize}
Thereby, we obtain 21 features from a single heart rate window, 18 features (i.e., three from the six axes of accelerometer and gyroscope) from a gait window, and 40 features from every breathing clip.

%\subsubsection{Feature Selection}\label{f_select}
\subsection{Feature Selection}\label{f_select}

To select the most influential features, we use the Sci-kit learn feature selection package and apply the three techniques: Correlation approach, Select from a Model (SelectFromModel), and Select the $K$ Best (SelectKBest). 
In each iteration of the leave-one-out training-testing, described in Section~\ref{train-test}, we select different sets features, which are very similar with changes in ordering.  
%Although each iteration had its own feature selection process the resulting features were very much inline with little variation, mostly just slight changes in the ordering of the selected features.\\
\textcolor{pink}{
%However, features selected across different iterations are very similar with changes in ordering. 
}

\begin{itemize}
    \item Correlation approach: We apply this approach to select the most influential heart rate and gait features. We find mean, variance, and skewness are the three uncorrelated features from both heart rate. Therefore, we remain with three heart rate features. \ignore{In a similar fashion we derived the 18 gait features. and 18 gait features obtained from the three axes of acceleration and gyroscope readings.} 
    \item Select from a Model (SelectFromModel): In this approach, we use the Random Forest Models. This feature selection approach provides a relative importance of features in percentages. %Figure~\ref{HG SelectFromModel} shows an example set of features selected using the SelectFromModel (with the explanation threshold, $p = 0.90$) approach. In the figure, green bars represent the features selected for modeling.
    \item Select the $K$ Best (SelectKBest): This is our last feature selection approach that also provides an importance score for each feature and based on that score we rank the features. Then, we try with different numbers of features, i.e., $K$, to find the best model performance. In this work, we find $K = 10$ performs the best. 
\end{itemize}

Tables~\ref{Best_Features} presents the list of best feature sets obtained from different biometrics using different selection approaches. 
Table~\ref{HGS_Features} presents the lists of selected features obtained using different selection approaches during one of the leave-one-out training-testing cases (Section~\ref{train-test}) considering heart rate, gait, and breathing features together.
\ignore{ Little variation between runs occurred and the table is displayed with the 1st iteration for simplicity.}

\begin{figure}[H]
    \centering
    \scalebox{.16}{\includegraphics{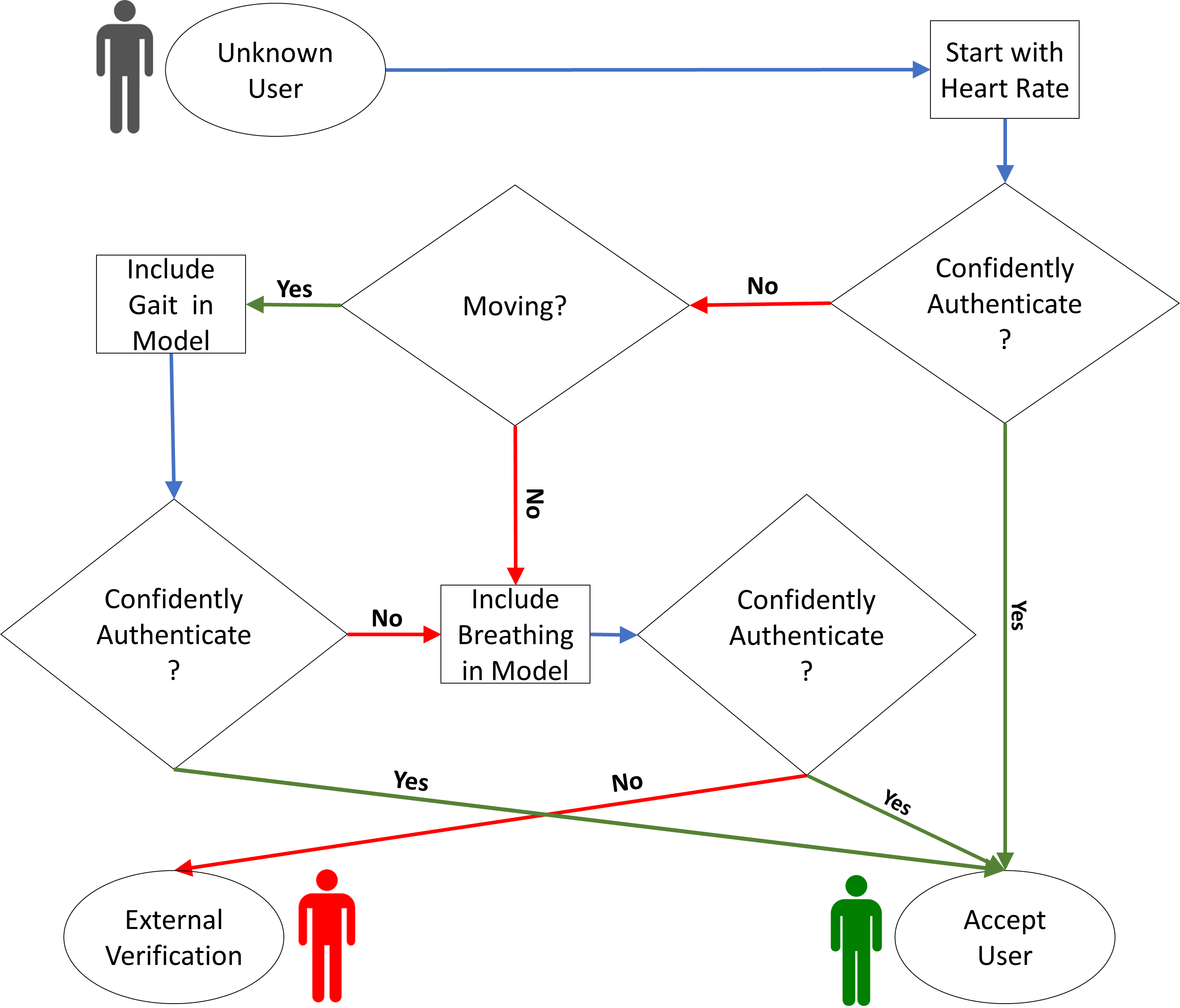}}
    \caption{\textcolor{black}{Proposed wearable device user authentication scheme}}
    \label{Bio Flow}
\end{figure}

\subsection{Methods}\label{methods}

In Figure \ref{Bio Flow}, we present an overview of our proposed implicit and continuous wearable-user authentication scheme utilizing various combinations of three biometrics (heart rate, gait, and breathing patterns) which are easily obtainable from most of the market wearables. 
The proposed system uses different collections of the three biometrics based on their availability and model confidence. 
%Depending on user's physiological state fusions of various combinations of biometrics will be used if needed to confidently authenticate the user.

We first try to authenticate a user based on the heart rate obtained from the photo-plethysmogram (PPG) sensor. If the system can authenticate the user with enough confidence, it allows the user to access the device. Otherwise, it checks the next authentication module that relies on other biometrics.

The authentication system first tries to check whether the user is moving using the on-device accelerometer and gyroscope data. If the user is moving, the system tries to authenticate the user based on gait and heart rate biometrics. If the system can authenticate the user with enough confidence, it allows the user to access the device. 

However, if the user does not move or the gait and heart rate-based module cannot authenticate the user, the system tries to combine breathing biometric collected from the on-device microphone. Based on the way the authentication system reaches the breathing module, it either combines breathing with only heart rate biometric or both heart rate and gait biometrics. 

Similar to the previous cases, the system in breathing module tries to authenticate the user. If the system can authenticate the user with enough confidence, it allows the user to access the device. Otherwise, the user's access to the device is revoked and require some sort of external verification, such as pin locks or passwords. 

Based on the various combinations of the three biometrics that we use in our authentication approach, we define the following models:
\begin{itemize}
\item
Heart rate data-driven model (HR model)
\item
Heart rate and gait data-driven model (HRG model)
\item
Heart rate and breathing data-driven model (HRB model)
\item
Heart rate, gait, and breathing data-driven model (HRGB model)
\end{itemize}

%% file: authentication.tex
\section{User Authentication}
\label{authentication}

Before presenting the detailed evaluation of our models, we first present training-testing set split and our modeling scheme, followed by list of performance measures and hyper-parameter optimization.

\subsection{Training-Testing Set}\label{train-test}

%% Will has to write down more about how he balanced the dataset, i.e., picked same number of windows/frames from each subject ... also you might talk a little bit about leave-one-out approach here
In our binary modeling, we try to distinguish a valid user (class 0) from the two impostors (class 1). To avoid overfitting, we consider at least 10 times more data-points, i.e., instances\ignore{feature windows} than the number of features. While training-testing, we follow the ``leave-one-out'' strategy, where we keep one instance for testing and use the rest of the $N-1$ instances for training with $N$ be the number of total instances from each class. 
We balance our training dataset, by considering the same $N-1$ number of instances from each class. Since our imposter class (class 1) consists of two person data, we pick $(N-1)/2$ instances from each imposter. Similarly, we balance our test sets. 
While adding the augmented data into train and test sets, we follow the same split that we consider on original data. For example, if a user has $N = 6$ breathing events and in our ``leave-one-out'' approach, we use the first five instances for training and the sixth instance for testing, then the breathing events that are generated from the first five events using different augmentation approaches are added to the training set. Similarly, events that are generated from the sixth breathing event are added to the test set. 
This way, we keep our training and testing instances mutually exclusive.

\subsection{Performance Measures}\label{perMeasures}
To evaluate the performance of different models we consider %\textcolor{blue}{
Accuracy (ACC), 
Root Mean Square Error (RMSE),
False Positive Rate (FPR), 
False Negative Rate (FNR),
$F_1$ score,
 and Area Under the Curve - Receiver Operating Characteristic (AUC-ROC). 
%}
Terminologies have their usual meaning in machine learning, when classifying a subject using a feature set~\cite{vhaduri2019multi,vhaduri2014estimating}. 
%\textcolor{blue}{
For an ideal system, it is desirable to have a lower RMSE, FPR and FNR, but a higher ACC, $F_1$ score, and AUC-ROC. 
\begin{table*}[!t]
\caption{The best HR, HRG, and HRG models with average and standard deviation of performance measures}
\label{Best_Results}
\centering
\begin{tabular}{l|l|c|c|c|c|c|c|c} %% Number of Columns 
\hline
Model & Classifier (parameters) & Feature count & ACC & RMSE & FPR & FNR & $F_1$ score & AUC-ROC \\
\hline 
HR  & SVM (poly. kernel, $d=2$, $C=1$) & 10 & 0.61 (0.18) & 0.62 (0.16) & 0.70 (0.20) & 0.08 (0.16) & 0.27 (0.09)& 0.54 (0.07)\\
\hline 
HRG  & RF (number of estimators, n = 500) & 13 & 0.84 (0.09) & 0.40 (0.01) & 0.26 (0.10) & 0.05 (0.03) & 0.86 (0.03) & 0.84 (0.04)\\
\hline
HRB  & $k$-NN ($k=6$, minkowski distance) & 10 & 0.91 (0.04) & 0.29 (0.04) & 0.17 (0.04) & 0.00 (0.00) & 0.92 (0.03) & 0.91 (0.02)\\
\hline
\end{tabular}
\end{table*}

\begin{table*}[!t]
\caption{The best HRGB models with average and standard deviation of performance measures}
\label{HGS_Results}
\centering
\begin{tabular}{l|c|c|c|c|c|c|c} %% Number of Columns 
\hline
Classifier (parameters) & feature count & ACC & RMSE & FPR & FNR & $F_1$ score & AUC-ROC \\
\hline
RF (n estimators = 450) & 23 & 0.91 (0.04) & 0.31 (0.00) & 0.19 (0.04) & 0.00 (0.00) & 0.91 (0.05) & 0.91 (0.02)\\
\hline
\rowcolor{lightgray} $k$-NN ($k=2$, minkowski distance) & 10 & 0.93 (0.06) & 0.27 (0.01) & 0.14 (0.07) & 0.00 (0.00) & 0.93 (0.03) & 0.93 (0.04)\\
\hline
NB & 10 & 0.91 (0.01) & 0.30 (0.00) & 0.18 (0.01) & 0.00 (0.00) & 0.92 (0.01) & 0.91 (0.01)\\
\hline
SVM (poly. kernel, $d=1$, $C=1$) & 10 & 0.92 (0.03) & 0.29 (0.01) & 0.17 (0.06) & 0.00 (0.00) & 0.92 (0.02) & 0.92 (0.03)\\
\hline
SVM (rbf kernel, $\gamma=0.01$, $C=1$) & 10 & 0.93 (0.03) & 0.27 (0.01) & 0.15 (0.06) & 0.00 (0.00) & 0.93 (0.03) & 0.93 (0.03)\\
\hline 
\end{tabular}
\end{table*}

%\subsection{User Authentication Models}

%\subsubsection{Finding Optimal Hyper-Parameter Sets}\label{param_opt}
\subsection{Hyper-Parameter Optimization}\label{param_opt}

We use the grid search package in the Sci-kit learn to find the optimal hyper-parameter sets. For each ``leave-one-out'' modeling, we \ignore{separately} perform the hyper-parameter optimization using various ranges of values.
From the different iterations of the ``leave-one-out'' approach with balanced mutually exclusive training-test sets, we obtain similar values for the hyper-parameters. 
%\textcolor{orange}{This is required to best show the performance of the balanced datasets as each iteration omits lots of imposter data.}
 \ignore{the displayed optimal hyper-parameters are those of the best ($F_1$ score) iteration for simplicity.}
In Tables~\ref{Best_Results} and~\ref{HGS_Results}, we present the set of optimal values obtained from different modeling approaches. 

%\subsubsection{Model Performance}\label{model_performance}
\subsection{Authentication Model Evaluation}\label{model_performance}

In Table~\ref{Best_Results}, we present the performance of the best models using various biometric combinations and different classifiers with their optimal parameter sets. 
%As the system process was discussed in Section \ref{methods} HR model is the initial metric used in order to attempt to authenticate the user.
%From Table \ref{Best_Results}, 
From the table, we observe that the best HR model (i.e., model that only uses heart rate data) can provide an average accuracy of $0.61 \pm 0.18$ and an average AUC-ROC $0.54 \pm 0.07$. 
As discussed previously in the Section~\ref{methods}, if the HR model cannot authenticate a user with enough confidence or fails to authenticate, we use additional gait biometric (i.e., HRG model).

In Table~\ref{Best_Results}, we observe that adding gait biometric (when available) with heart rate, all measures improve. In case of the best HRG model (i.e., model that uses heart rate and gait biometrics), ACC increased by 38\%, $F_1$ score increased by 218\%, and AUC-ROC increased by 56\% compared to the best HR model. 
%The accuracy jumps to $0.84\pm0.09$. 
The FPR also improves (i.e., drops) from $0.70 \pm 0.20$ to $0.26\pm0.10$. While gait data is only available while a user is moving, its inclusion \ignore{is always available, but}can significantly boost the authentication performance, compared to a model that uses only less accurate minute-level heart rate data. 
%and this is exclusive to the lessening confidence threshold requirements from HR model.

In Table~\ref{Best_Results}, we observe that the HRB model (i.e., model that uses heart rate and breathing biometrics) achieves a better performance compared to the HRG model. We achieve a 35\% drop in the FPR while comparing the HRB  with the HRG model. Additionally, we observe $\approx$ 8\% increases, while comparing the ACC, $F_1$ score, and AUC-ROC of the HRB model with the HRG model. 
While comparing the HRB model to the HR model, we observe a huge performance improvement. Compared to the HR model, the HRB model performs better with a $F_1$ score (an increase of 241\%) and AUC-ROC (an increase of 68\%) with a high accuracy of $0.91 \pm 0.02$.

Finally, in Table~\ref{HGS_Results}, we present the performance comparison among different classification models and their optimal parameter sets using the heart rate, gait, and breathing biometrics together. 
In the table, we observe a modest performance improvement, i.e., an accuracy increase of accuracy by 2\% (0.91 to 0.93), while comparing the HRGB model with the HRB model.  
Similar improvements are observed in case of other performance metrics. 
However, in case of the HRGB model we obtain a simpler and faster classifier, i.e., the $k$-NN with $k = 2$ (number of neighbors needed for a classification), compared to the HRB model that uses the $k$-NN model with $k = 6$. Therefore, considering the implementation endowment, i.e., a wearable, that has energy constraints, an authentication model that uses heart rate, gait, and breathing biometrics and the $k$-NN classifier with $k = 2$ can be a better choice. 

%On top of that as seen between Figures \ref{Error Rate HR Sound} and \ref{Error Rate All} a lower confidence threshold is needed to drop the FPR.

\begin{figure}[ht]
    \centering
    \scalebox{.53}{\includegraphics{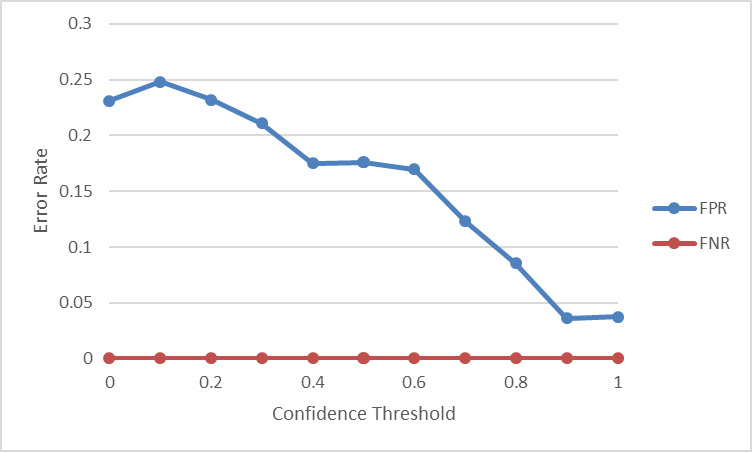}}
    \caption{\textcolor{black}{The change of FPR and FNR with varying confidence thresholds (HRB model with the $k$-NN classifier)}}
    \label{Error_HRB}
    \end{figure}

\begin{figure}[ht]
    \centering
    \scalebox{.53}{\includegraphics{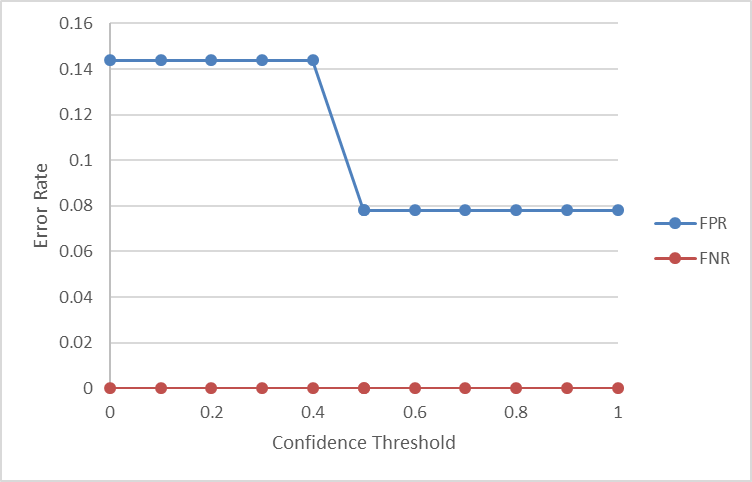}}
    \caption{\textcolor{black}{The change of FPR and FNR with varying confidence thresholds (HRGB model with the $k$-NN classifier)}}
    \label{Error_HRGB}
    \end{figure}

%\subsubsection{Accessing Error Rate}\label{accessing_error_rate}
%\subsubsection{Error Analysis}\label{accessing_error_rate}
\subsection{Error Analysis}\label{accessing_error_rate}

As discussed previously in Section~\ref{methods}, the authentication system allows a user to access the device only when it can validate the user with enough confidence. In this section, we present an analysis on how our system performs with the change of confidence level, i.e., threshold. In case of an ideal system, it is desired to have a lower FPR and FNR. 
In Figures~\ref{Error_HRB} and~\ref{Error_HRGB}, we present our analysis of error rates (FPR and FNR) with varying confidence thresholds. In Figure~\ref{Error_HRB}, we observe that FPR sharply drops with the increase of thresholds from 0.1 to 0.9 and the FPR drops below 0.05 at confidence threshold 0.9. 
However, in Figure~\ref{Error_HRGB}, we observe a sharp drop between threshold values of 0.4 and 0.5. The FPR drops from 0.14 to 0.08 during this increase of threshold and FPR remains steady before and after that change in threshold. 
Though the FPR in case of HRGB model remains a little bit high compared to the HRB model, but the HRGB model needs a smaller confidence threshold of 0.5 to achieve this FPR and at this 0.5 confidence threshold HRGB model can drop the FPR $\approx$ 54\% compared to the HRB model (i.e., FPR of 0.08 versus 0.175). Therefore, it is desirable to use an authentication model that uses heart rate, gait, and breathing biometrics with a low confidence threshold of 0.5.

%% file: discussion.tex
\section{Limitations, Discussion, and Conclusions}

Our work tested the feasibility of authenticating users implicitly and continuously based on the three separate biometrics, i.e., heart rate, gait, and breathing, that are easily obtainable in most of the market wearables, which are usually not equipped with powerful sensors, such as cameras or fingerprint scanners.
To the best of our knowledge, this is the first work that attempts to authenticate a wearable device user without any explicit user burden using three different soft-biometrics in a more case-based approach, i.e., availability of data. 
Through our detailed analysis we show that we can authenticate a user with an average accuracy of $0.93\pm0.06$, $F_1$ score of $0.93\pm0.03$, and AUC-ROC of $0.93\pm0.04$, with a less than 0.08 FPR at 50\% confidence threshold using three biometrics together.
This shows the promise for developing an implicit and continuous authentication system for the market wearables to secure our valuable information as well as to create a safe gateway to access various services and devices.  

This work has some limitations, which we plan to address in the future. First, we have limited number of audio breathing clips. However, we increase the data volume using standard audio augmentation approaches. Second, in this feasibility work, we use a set of three subjects. We perform a leave-one-out validation approach to deal with this limitation and our achieved performance measures show a promise to further investigate this with a large-scale extended period study. Third, we use different datasets, which could affect the performance. However, we use three independent biometrics and perform correlation analysis to select uncorrelated features; thereby, our results potentially shows a baseline performance, which could further be improved by using the three biometrics from the same subject since that could more uniquely identify a user compared to our case. 
Before mass deployment, there needs a study for an extended period with several subjects to incorporate user variability and behavior changes over time. 
Finally, more advanced modeling techniques such as deep learning models (recurrent neural networks or convolutional neural networks) may further improve the accuracy of the models, but that will require to off load data from the wearable to server, which can lead to additional challenges to secure; therefore, our approach of using lighter machine learning models, such as the $k$-NN with $k = 2$, have a higher scope to implement on the wearables.